\documentstyle[prl,aps]{revtex}
\begin{document}
\title{Beta-relaxation of non-polymeric liquids close to the
glass transition}
\twocolumn
\author{Niels Boye Olsen, Tage Christensen, and Jeppe C. Dyre}
\address{Department of Mathematics and Physics (IMFUFA), 
Roskilde University, Postbox 260, DK-4000 Roskilde, Denmark}
\date{\today}
\maketitle{}

\begin{abstract}
Dielectric beta-relaxation in a pyridine-toluene solution is
studied close to the glass transition.
In the equilibrium liquid state the beta loss peak frequency is
not Arrhenius (as in the glass) but virtually 
temperature-independent, while the maximum loss is strongly
temperature-dependent.
Both loss peak frequency and maximum loss exhibit thermal
hysteresis.
A new annealing-state independent parameter involving loss and
loss peak frequency is identified.
This parameter has a simple Arrhenius temperature-dependence
and is unaffected by the glass transition.
\end{abstract}


Viscous liquids are characterized by relaxation times that
increase strongly upon cooling towards the glass transition
\cite{har76,bra85,ang91,edi96}.
The relaxation time of molecular rotation is monitored by
dielectric relaxation experiments probing the linear response
to a periodic external electric field
\cite{mcc67,kiv77,wil78,bot78,bag91,hav95,jon96}.
The dominant relaxation process is referred to as the
alpha-process.
For most viscous liquids, upon cooling the alpha-peak
bifurcates just before the glass transition and an additional
minor loss peak appears at higher frequencies 
\cite{ros93,don96,ros97,wil73}.
This is traditionally referred to as beta-relaxation
\cite{mcc67,joh76,wil79,joh85,kud99} (now sometimes termed
Johari-Goldstein beta-relaxation to distinguish it from the
mode-coupling theory's ``cage-rattling'' at much higher
frequencies).
Beta-relaxation has also been observed in mechanical
\cite{ber83,joh87} and thermal \cite{fuj95} relaxation
experiments.
Here, we limit ourselves to dielectric beta-relaxation.
Our purpose is to show that the conventional view that
beta-relaxation is unaffected by the glass transition is not
confirmed by experiments on non-polymeric liquids.
We have not studied beta-relaxation in polymers, but believe
based on the literature that beta-relaxation in polymers is
probably {\em not} affected by the glass transition.

Beta-relaxation was first seen in polymers, where it was
attributed to side-chain motion \cite{mcc67}.
In 1970 Johari and Goldstein found beta-relaxation in a number
viscous liquids of rigid molecules and conjectured that
beta-relaxation should be considered as ``a characteristic 
property of the liquid in or near the glassy state'' 
\cite{joh70a,joh70b}.
However, for some glass-forming liquids (e.g., glycerol) no 
beta-relaxation has been observed.
Today, viscous liquids are sometimes classified according
to whether
or not they exhibit beta-relaxation \cite{kud99}, although
there are recent intriguing speculations that beta-relaxation
indeed {\it is} universal with the beta-peak sometimes hiding
under the alpha-peak \cite{nga98}. 

There is no general agreement about the cause of 
beta-relaxation \cite{kud99,han97b}.
It is unknown whether every molecule contributes to the
relaxation \cite{wil71,han97} or only those within ``islands of
mobility'' \cite{joh76,gol69}.
Similarly, it is not known whether small angle jumps
\cite{kau90} or large angle jumps \cite{arb96} are responsible
for beta-relaxation.
Of course, a possible explanation of these disagreements is
that beta-relaxation is non-universal \cite{kah97,wag98}.

As traditionally reported in the literature (see, e.g.,
\cite{joh76}), beta-relaxation is characterized by a broad loss
peak with Arrhenius temperature-dependent loss peak frequency
and only weakly temperature-dependent maximum loss.
In this picture, which is mainly based on measurements in the
glassy phase, the glass transition has no effect on the
temperature-dependence of the beta loss peak frequency.
In our opinion, it is unlikely that the temperature-dependence
of the loss peak frequency is unaffected by the glass
transition, considering the well-known fact that the {\it
strength} of the beta process in the glassy state decreases
during annealing \cite{joh71,joh82} (in some cases to below
resolution limit \cite{han97}).
Actually, no detailed investigations of beta-relaxation in the
equilibrium liquid phase of non-polymeric liquids have been
carried out. 
This may be because studying beta-relaxation above the glass
transition temperature $T_g$ is difficult since there is only a
tiny temperature-window (if any) where alpha- and
beta-relaxations are well-separated.

Motivated by the above arguments, one of us recently
investigated beta-relaxation in sorbitol and found that the
temperature-dependence of both loss peak frequency and loss
magnitude in the equilibrium liquid state is indeed different
from what is found in the glassy state \cite{ols98}.
This result was obtained on a system which - like most others -
has a beta-relaxation that in the equilibrium liquid phase is
not very well separated from the alpha-relaxation.
The sorbitol results were mainly based on annealing experiments
below $T_g$ and the results were to some extent inconclusive.
Below, we present data for beta-relaxation in a liquid with a
strong beta-peak which is well separated from the alpha-peak in
a range of temperatures above $T_g$. 
The liquid is a 71\%/29\% mixture of pyridine and toluene, a
system first studied by Johari \cite{joh82}.
Toluene molecules have only a small dipole moment, so
dielectric spectra mainly reflect motion of the pyridine
molecules acting as probes of the overall dynamics of the
solution \cite{wil72}.

The dielectric measuring cell used is a 22-layer gold-plated
capacitor with empty capacitance 68 pF (layer distance 0.1 mm).
The dielectric constant was measured over 9 decades of
frequency using standard equipment:
From 100 Hz to 1 MHz a HP4284A precision LCR meter was used,
from 1 mHz to 100 Hz a HP3458A multimeter in conjunction with a
Keithley 5 MHz, 12-bit, arbitrary waveform generator was used.
The dielectric loss was determined with at precision better
than $10^{-4}$ in the whole frequency-range.
The measurements were carried out in a cryostat designed for
long time annealing experiments, keeping temperature variations
below 5 mK.

Figure 1a shows the dielectric loss at 125K, 126K, and 127K.
The alpha- and beta-peaks are quite well separated.
Despite this a procedure is needed to eliminate the alpha-tail
influence on the beta-peak.
From Fig. 1a we find that the alpha-peak follows a
high-frequency power-law decay with exponent -0.47.
In order to arrive at the ``true'' beta-peak this alpha-tail
was subtracted by applying the following procedure:
At each temperature the magnitude of the subtraction was 
uniquely determined by requiring that the beta-peak follows
a low-frequency power-law.
We used a power-law fit \cite{hav95,jon96} because a Gaussian,
as sometimes used to fit beta-peaks \cite{wu92,kud97}, cannot
fit our data.
This way to eliminate the alpha-contribution involves the
following assumptions:
1) The dielectric spectrum is a simple sum of alpha- and 
beta-relaxation and not, e.g., a Williams-convolution
\cite{wil79,ber98}; 
2) In the relatively narrow temperature-interval under study
the alpha-tail's power-law decay has an exponent which is
temperature-independent.
Figure 1b shows eight normalized beta-peaks (119K-126K) after
subtraction of alpha-tails.
The figure shows that the subtraction procedure is consistent:
The ``corrected'' beta-peaks do follow a low-frequency
power-law to a good approximation.

Figure 2a shows beta loss peak frequency $f_{max}$ ($\Box$)
and maximum loss $\epsilon_{max}$ ($\Diamond$) as function of
inverse temperature for a cooling taking the equilibrium liquid
into the glassy state at a rate of 1 K/h.
The system was cooled in steps of 0.5 K.
Dielectric loss was measured after annealing 30 minutes at
constant temperature, immediately before cooling another 0.5 K.
At high temperatures - in the equilibrium liquid state - the
loss peak frequency is almost temperature-independent
\cite{wilnote} while the loss decreases sharply during the
cooling \cite{kud98}.
At low temperatures, the well-known Arrhenius 
temperature-dependence of loss peak frequency is observed and
the maximum loss is much less temperature-dependent than in the
liquid.

Figure 2b shows beta loss peak frequency $f_{max}$ and maximum
loss $\epsilon_{max}$ as function of inverse temperature during
a cooling through the glass transition followed by a subsequent
faster reheating.
Starting in the equilibrium liquid state the sample was cooled
in steps of 0.5 K with measurements carried out after annealing
30 minutes at each temperature.
The cooling continued until 119 K was reached.
The sample was then heated in steps of 1.0 K every 30 minutes.
The figure shows hysteresis like that found for all other
quantities changing their temperature-dependence at the glass
transition.
The figure also shows ($\triangle$) the temperature-dependence
during both cooling and subsequent reheating of the following 
quantity

\begin{equation}\label{1}
X\ =\ f_{max} \left(\epsilon_{max}\right)^{\gamma}\,.
\end{equation}
Here, $\gamma = 1.19$ is an empirical exponent.
There is just one curve marking the temperature-dependence of
$X$, showing that $X$ exhibits no thermal hysteresis.
In particular, $X$ is independent of annealing-state.
Surprisingly, $X$ has an Arrhenius temperature-dependence.

We have found similar behavior for sorbitol, DBP, DEP, DMP, and
PPG, although these liquids have beta-peaks that are less 
well-separated from the alpha-peak in the liquid state.
The exponent $\gamma$ is non-universal (varying from 1 to 2).
We have no model for these findings.
Speculating, we note that the case $\gamma=1$ may be modelled
by an asymmetric two-level system:
If the large barrier is temperature-independent $X$ is
Arrhenius \cite{dyr}.
However, in order to explain the findings of Fig. 2, a rather
peculiar temperature-dependence of the highest of the two
energy minima must be assumed \cite{dyr}.

In conclusion, we have shown that the loss peak frequency of
beta relaxation in a pyridine-toluene solution is almost
temperature-independent above $T_g$ where, on the other
hand, the maximum loss is strongly temperature-dependent.
Thus, beta relaxation in the equilibrium liquid state has
characteristics that are opposite of those found in the glassy
state, where loss peak frequency is Arrhenius and maximum
loss is only weakly temperature-dependent.
It has furthermore been shown that the quantity $X$ of Eq.\
(\ref{1}) is Arrhenius; in particular $X$ exhibits no thermal
hysteresis around $T_g$.
These results contradict the traditional view of
beta-relaxation as being unaffected by the glass transition and
show the need for further experimental as well as theoretical
work in this field.
The recent surprising findings by Wagner and Richert that
liquids like o-terphenyl and salol, previously believed to have
no beta-relaxation, do exhibit beta-relaxation deep in the
glassy state after very fast quenchings \cite{wag98,ric98}
emphasize the need for further work in this field.

\acknowledgements
This work was supported in part by the Danish Natural Science
Research Council.

\begin{figure}\caption[lgt]

{\it Fig. 1a:}

Log-log plot of dielectric loss as function of frequency for a
71\%/29\% mixture of pyridine in toluene at 125 K, 126 K, and
127 K (16 measured points per decade, no smoothing applied).
The large low-frequency peak is the alpha-relaxation process,
the small high-frequency peak is beta-relaxation.
The fact that the entire alpha-peak is visible at 126 K and 127
K signals that these measurements were taken in the equilibrium
liquid phase, i.e., above the glass transition.
In contrast to most other viscous liquids the system is
characterized by a clearly visible beta-peak in the equilibrium
liquid.
The alpha-relaxation is characterized by a power-law tail,
proportional to $f^{-0.47}$.

{\it Fig. 1b:}

Log-log plot of the normalized beta-peak at eight temperatures
(T=119-126K) after subtraction of the high-frequency alpha-tail
$\propto f^{-0.47}$.
The magnitude of the subtraction was determined uniquely from
requiring that the beta-peak follows a low-frequency power-law 
(leading to the power-law $\propto f^{0.45}$ at all
temperatures).
The figure shows that the assumptions behind this procedure are
consistent, the assumptions being:
1) simple additivity of alpha- and beta-relaxation, 2)
alpha-tail given by a power-law decay with a 
temperature-independent exponent, and 3) beta-relaxation
at low-frequencies following a power-law.
Deviations are found at low frequencies for the highest
temperatures (here the alpha-peak is so close to the
beta-peak that the power-law alpha-tail subtraction
overestimates the alpha-contribution).

\end{figure}

\begin{figure}\caption[lgt] 

{\it Fig. 2a:}

Logarithm of beta loss peak frequency ($\Box$) and maximum
loss ($\Diamond$) as function of inverse temperature for a
cooling from 126.5 K to 119.0 K (raw data before subtraction of
alpha-tail marked by dots).
The system was cooled in steps of 0.5 K with dielectric
measurements carried out after annealing for 30 minutes at each
temperature immediately before stepping down another 0.5 K.
The figure shows a clear change of behavior at the glass
transition which takes place around x=8.1, corresponding to 
$T_g\cong$123.5 K.
Above $T_g$ (in the equilibrium liquid) the loss peak frequency
is almost temperature-independent while maximum loss is
Arrhenius, below $T_g$ the opposite is the case.

{\it Fig. 2b:}

Logarithm of beta loss peak frequency ($\Box$) and maximum loss
($\Diamond$) for a cooling from 126.5 K to 119.0 K at 1 K/h as
in Fig. 2a and reheating at the double rate.
Both quantities exhibit hysteresis as expected for quantities
that change their temperature-dependence at the glass
transition.
The symbol $\triangle$ marks the quantity
$X=f_{max}\left(\epsilon_{max}\right)^{1.19}$ for both cooling
and subsequent reheating.
$X$ exhibits no hysteresis, is insensitive to the glass
transition, and is Arrhenius temperature-dependent.

\end{figure}

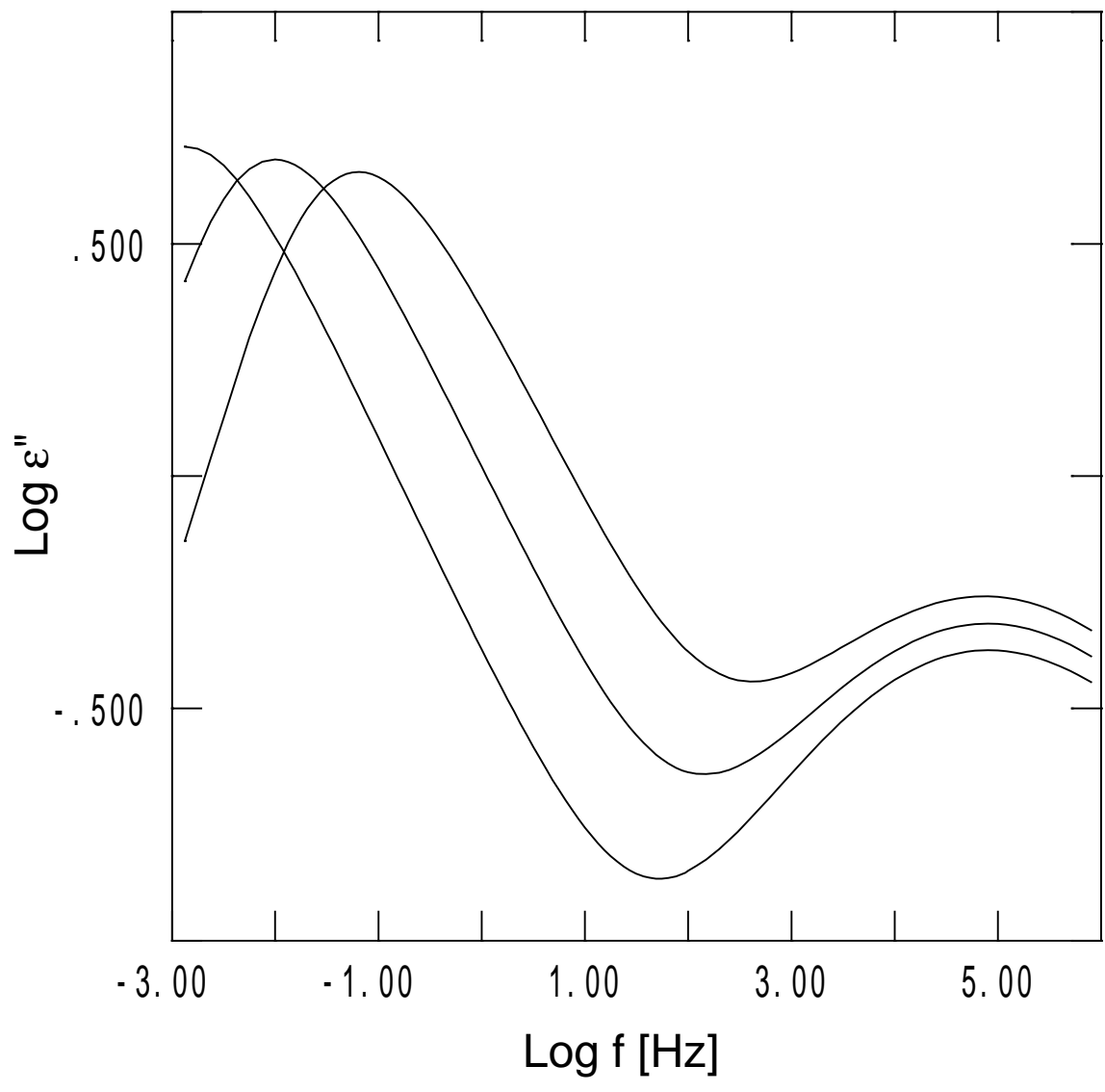
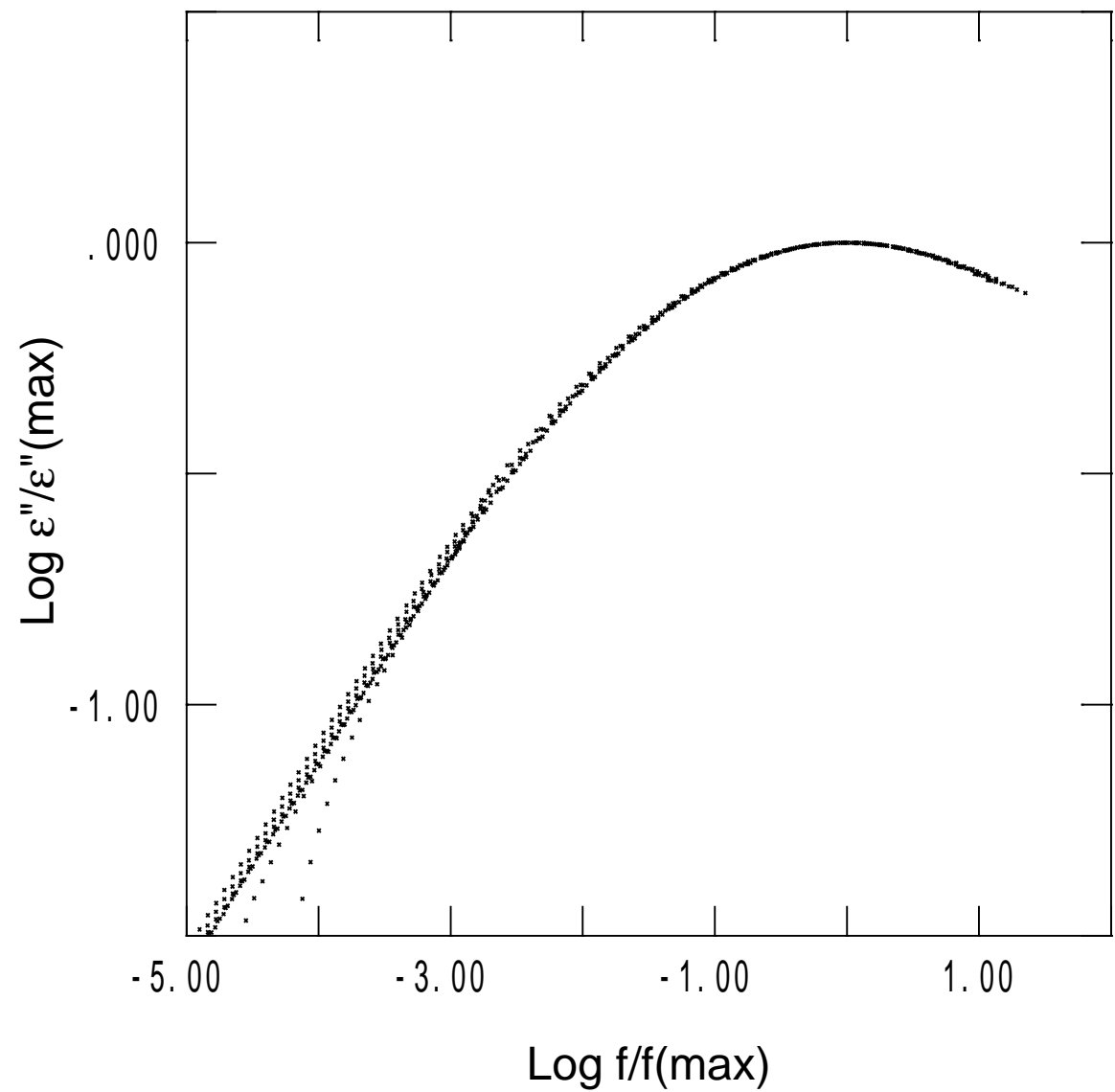
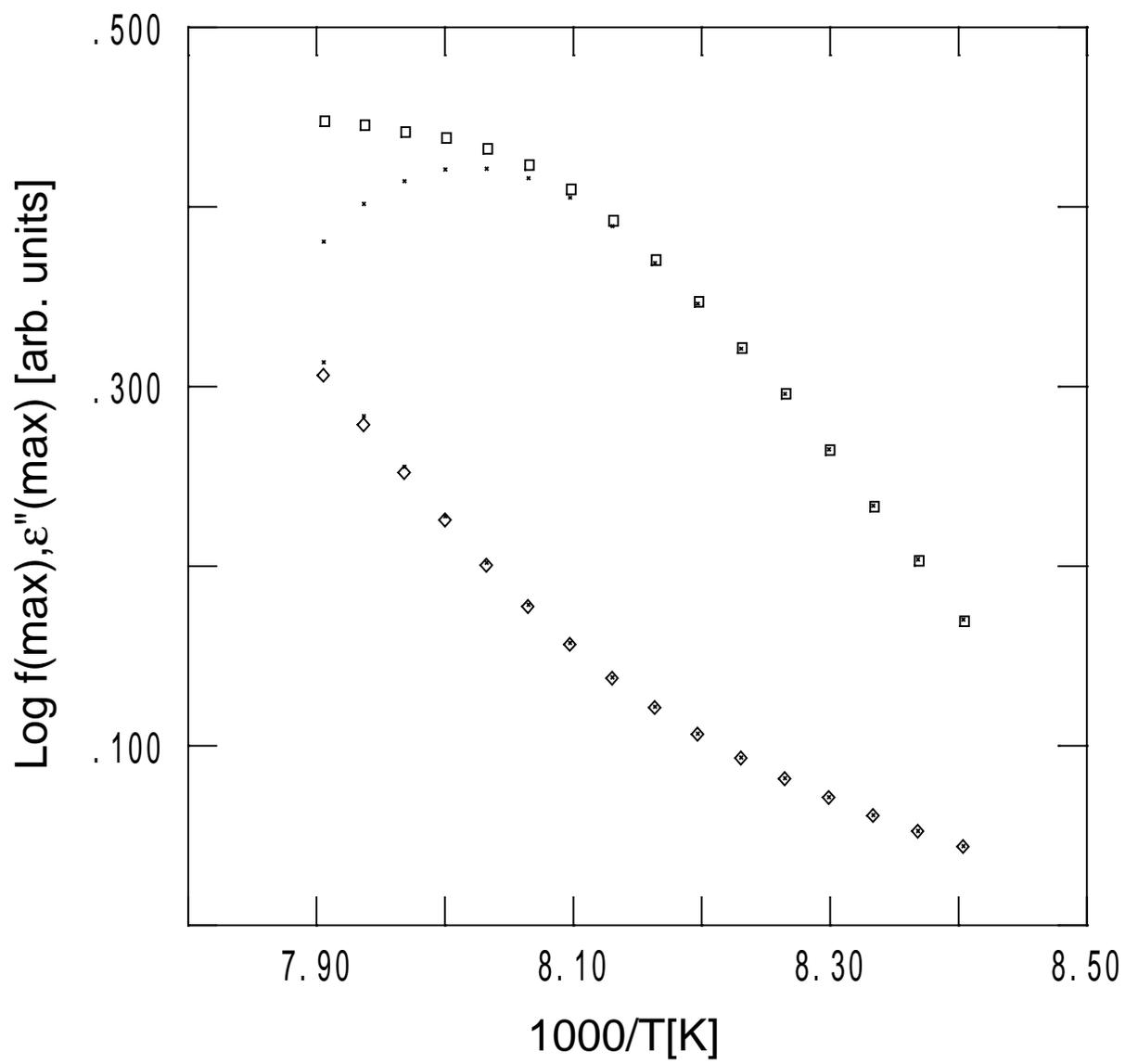
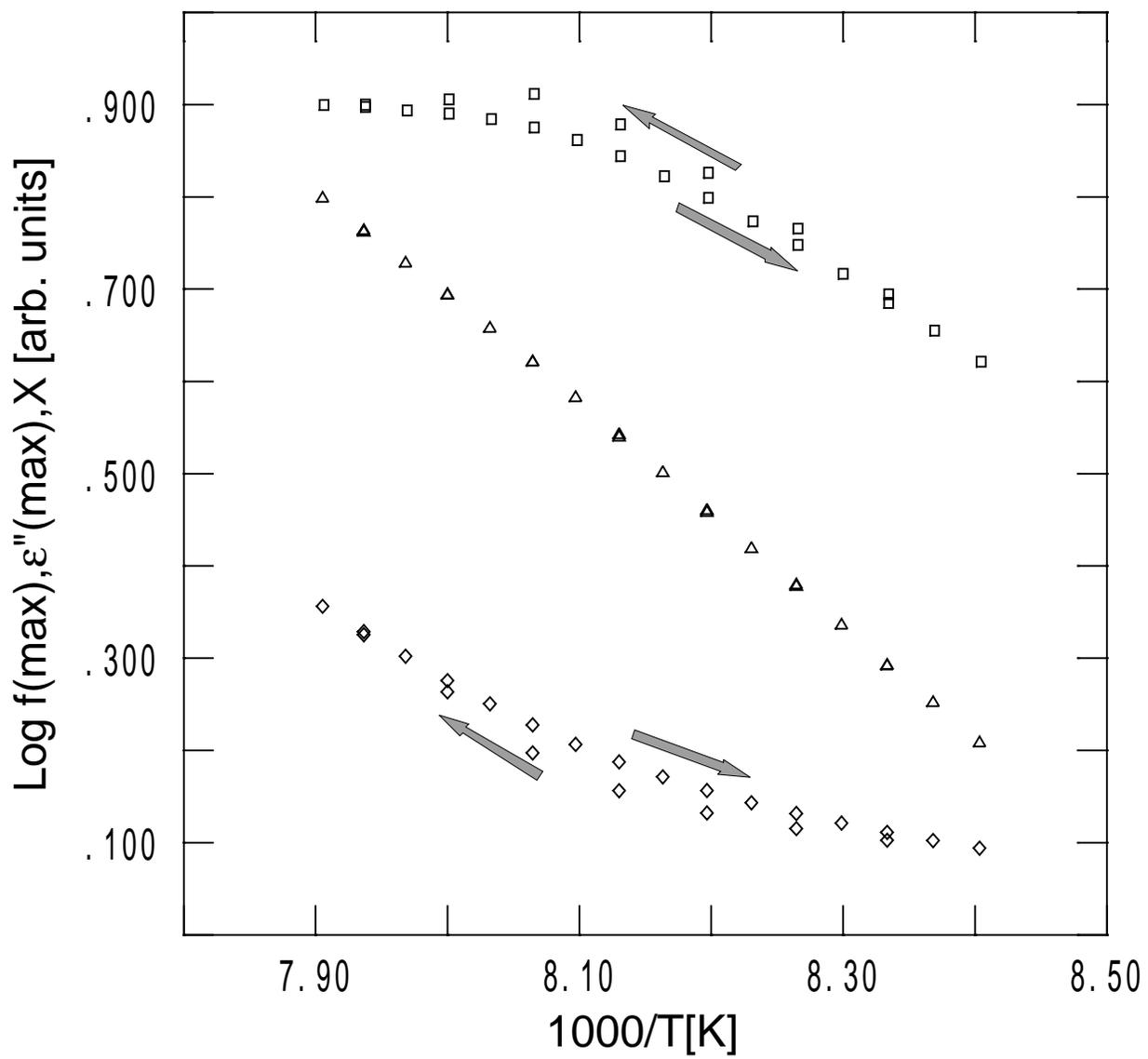
\end{document}